\UseRawInputEncoding
\documentclass[10pt, conference]{IEEEtran}
\usepackage{cite}
\usepackage{amsmath,amssymb,amsfonts}
\usepackage{graphicx}
\usepackage{textcomp}
\usepackage{xcolor}
\usepackage{multirow}
\usepackage{booktabs}
\usepackage{mdframed}
\usepackage{xspace}
\usepackage{caption}
\usepackage{hyperref}
\hypersetup{hidelinks}

\newcommand{\kg}{{WebAppArchKG}\xspace}
\newcommand{\app}{{WebDesignIter}\xspace}

\begin{document}

\title{WebDesignIter: Co-Evolving Design Knowledge for Repository-Level Front-End Code Generation}






\author{
\IEEEauthorblockN{Zheng Pei}
\IEEEauthorblockA{%
  \textit{Sun Yat-sen University}\\
  \textit{Zhuhai Key Laboratory of Trusted Large Language Models}\\
  Zhuhai, China\\
  zhengpei516@gmail.com
}
\and
\IEEEauthorblockN{Mingwei Liu}
\IEEEauthorblockA{%
  \textit{School of Software Engineering, Sun Yat-sen University}\\
  \textit{Zhuhai Key Laboratory of Trusted Large Language Models}\\
  Zhuhai, China\\
  liumw26@mail.sysu.edu.cn
}
\and
\IEEEauthorblockN{Zhenxi Chen}
\IEEEauthorblockA{%
  \textit{Sun Yat-sen University}\\
  \textit{Zhuhai Key Laboratory of Trusted Large Language Models}\\
  Zhuhai, China\\
  chenzhx236@mail2.sysu.edu.cn
}
\and
\IEEEauthorblockN{Zihao Wang}
\IEEEauthorblockA{%
  \textit{Sun Yat-sen University}\\
  \textit{Zhuhai Key Laboratory of Trusted Large Language Models}\\
  Zhuhai, China\\
  wangzh778@mail2.sysu.edu.cn
}
\and
\IEEEauthorblockN{Yanlin Wang}
\IEEEauthorblockA{%
  \textit{Sun Yat-sen University}\\
  \textit{Zhuhai Key Laboratory of Trusted Large Language Models}\\
  Zhuhai, China\\
  wangylin36@mail.sysu.edu.cn
}
}

\maketitle

\begin{abstract}
Front-end development accumulates change after change at the repository level, weaving complex cross-file dependencies that current LLM coding agents — tuned for single-shot tasks — cannot reliably track across multiple iterations, leading to functional regressions and code that resists maintenance. We argue the missing piece is design knowledge: architectural principles, module responsibilities, and structural constraints that developers lean on to keep code readable, maintainable, and evolvable as a system scales. To operationalize this, we propose \app, a framework built around a persistent knowledge graph (\kg) that fuses repository structure with design knowledge and keeps both in sync across development cycles. \app works in two stages: design-informed planning pulls historical context and architectural overviews from \kg to produce an implementation plan with corresponding test scripts, and design-aware generation executes that plan through targeted diff-based patches, validated by sandbox execution and automatic syntax repair.

On Web-Bench, \app delivers an average Pass@2 gain of 9.55 percentage points across nine foundation models over existing baselines. More importantly, \app outperforms every general-purpose coding agent — Claude Code, OpenHands, SWE-Agent, Codex CLI — on every model configuration, posting the highest Pass@1 and Pass@2 while consuming 25–30× fewer input tokens. Ablation singles out design knowledge as the most impactful component: stripping it drops Pass@1 by 11.40 percentage points, a degradation far larger than removing code-graph retrieval, patch-based generation, or sandbox verification, confirming that design knowledge provides a fundamentally more efficient and reliable path to repository-level code generation.
\end{abstract}



Front-end development is critical in modern software systems, as it directly impacts user interaction quality and overall usability. Compared to back-end development, it involves a heterogeneous technology stack and strongly component-oriented structures~\cite{stefanova2024exploring}, encompassing multiple file types (e.g., HTML, CSS, JavaScript/TypeScript, and configuration files) and complex cross-file and cross-component dependencies~\cite{kurant_static_2025}. In practice, front-end development evolves incrementally at the repository level~\cite{lehman2005programs}, where new features are added while preserving existing functionality. This tightly coupled, continuously evolving paradigm challenges repository-level code generation and maintenance, particularly regarding semantic understanding, contextual construction, and regression risk.

Large language models (LLMs) have recently become primary assistive tools in front-end development and have demonstrated strong performance across a wide range of software engineering tasks, including code generation, completion, translation, security analysis, and automated repair~\cite{lin2403soen, yang2024morepair, chen2025llms, hong2025automatically, li2024fine, liu2024stall+, williams2026empiricalsustainabilityaspectssoftware, 10.1145/3728978}. Building on these advances, agent-based approaches~\cite{otoum2026methods} and the Vibe Coding paradigm~\cite{sapkota2025vibecodingvsagentic} have gained popularity. However, \textbf{these methods typically optimize for single-task completion and struggle in long-term, continuously evolving repositories, where they often induce functional degradation or regressions of previously implemented features}.

Most existing front-end–oriented research further focuses on UI2Code paradigms~\cite{wu2025mllm, 2019-Stocco-Proweb, designcoder, laurencon2024unlocking, wan2024automatically}, which translate visual interface designs into code. While effective for prototyping, such approaches often produce simplified implementations with limited interactivity and fail to capture the complexity of real-world front-end repositories that evolve incrementally over time~\cite{webbench}. As a result, one-shot UI-to-code generation remains insufficient for realistic front-end engineering workflows.

\begin{figure}[htbp]
  \centering
  \includegraphics[width=1.0\columnwidth]{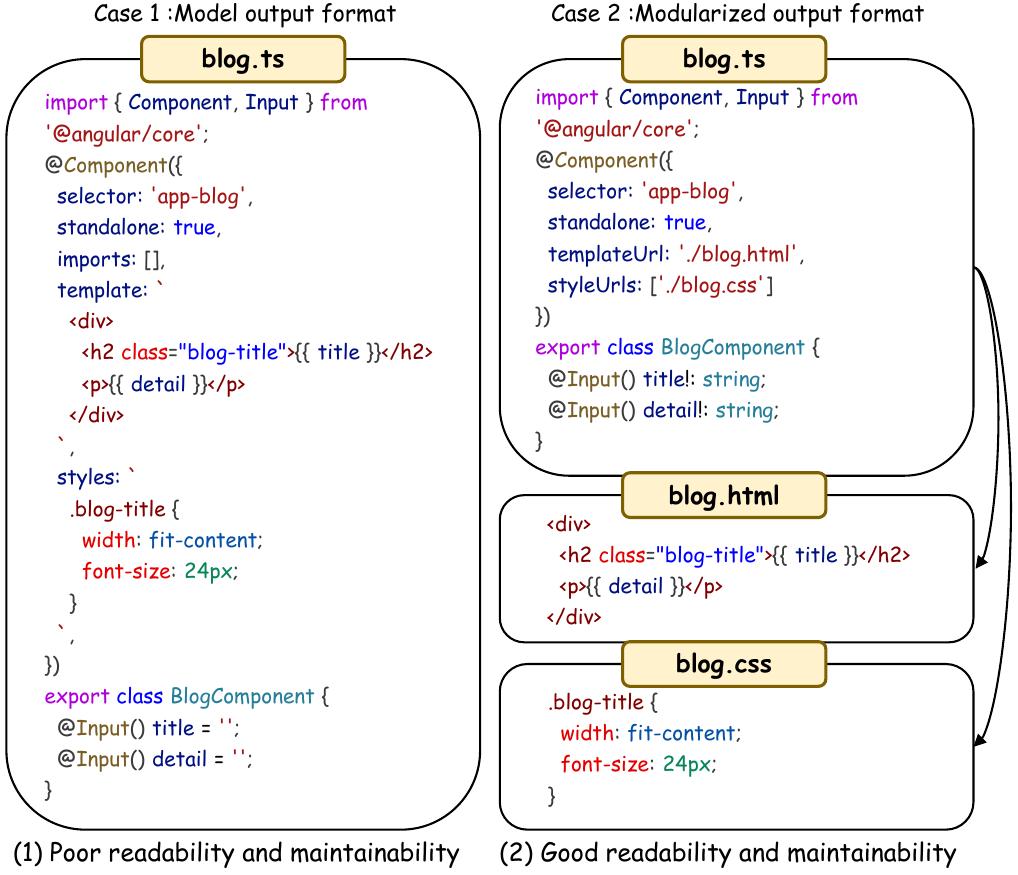}
  \caption{The impact of modularized output formats and model output formats on readability and maintainability}
  \label{fig:code_file}
\end{figure}

A key missing component in existing approaches is design knowledge---the structured understanding of a software system's architecture, encompassing component responsibilities, module boundaries, dependency constraints, and the design rationale behind structural decisions. \textbf{In real-world front-end development, such knowledge is essential for long-term readability, maintainability, and evolution~\cite{design}}. Poor design practices easily lead to ``spaghetti code'', such as mixing presentation, styling, and business logic within the same file, as illustrated in Figure~\ref{fig:code_file}. This significantly increases the cost of understanding and modifying code in subsequent iterations while reducing code readability and maintainability. Moreover, such low-quality designs introduce redundant and irrelevant information, inflating the contextual burden on LLMs and further degrading generation quality.

Recent benchmarks, such as Web-Bench~\cite{webbench}, target incremental, multi-task, and repository-level front-end development settings and systematically expose the complexity and challenges of these scenarios by simulating realistic software development workflows. On this benchmark, Web-Agent~\cite{webbench} achieves only 42.40\% Pass@2 and an even lower 26.40\% Pass@1, indicating limited overall performance and suggesting that the majority of tasks remain difficult to complete successfully. These results further demonstrate that even state-of-the-art agent-based approaches struggle in such environments, largely due to their limited ability to integrate design knowledge and reason about system-level dependencies, which often leads to loosely structured and poorly maintainable code.

Addressing these challenges requires overcoming three fundamental obstacles: (1) design knowledge and repository information must be persistently maintained and co-evolved across iterations to preserve global consistency; (2) automated refactoring of poorly structured code is necessary to sustain long-term readability and reduce contextual noise; and (3) precise cross-file and cross-component context construction is essential for effective repository-level generation in heterogeneous front-end systems.

To this end, we propose \app, a design-knowledge-driven framework for incremental front-end repository-level code generation. \app maintains a persistent repository-level knowledge graph (\kg) that integrates code structure and design knowledge, supporting a two-stage generation pipeline: the Design-informed Planning stage generates an Implementation Plan driven by Design Knowledge, followed by the Design-aware Generation stage, which produces precise code modifications. By explicitly embedding design knowledge into the generation process, applying code changes, fixing syntax errors, refactoring spaghetti-like files to enhance readability and maintainability, and validating correctness in a sandbox before finally updating \kg, \app enables incremental, maintainable, and verifiable front-end development.

Experiments on Web-Bench~\cite{webbench} show that \app achieves state-of-the-art performance, with an average improvement of \textbf{7.98\%} in Pass@1 and \textbf{9.55\%} in Pass@2 over existing baselines. \app also outperforms general-purpose coding agents---including Claude Code, OpenHands, SWE-Agent, and Codex CLI---across every model configuration, attaining the highest Pass@1 and Pass@2 while consuming 25--30$\times$ fewer input tokens. Ablation studies confirm the critical role of design knowledge, whose removal causes a 11.40\% drop in Pass@1. Maintainability analysis further shows that \app reduces average file length and repository complexity, enhancing code readability.

In summary, our contributions are threefold:
\begin{itemize}
    \item \textbf{Design Knowledge-Guided Front-End Repository-Level Code Generation:} We integrate design knowledge into front-end repository-level code generation. High-Level Design produces an abstract implementation plan, while Low-Level Design uses this plan to generate precise, functionally complete code diff patches.

    \item \textbf{Construction of \kg:} Our knowledge graph maintains design knowledge, historical design, and repository structure, supports update strategies and cross-file component associations, and automatically refactors poorly structured code to improve readability and maintainability.

    \item \textbf{Diff Patch and Syntax Check Mechanisms:} We employ incremental diff patches to avoid regressions caused by full-file regeneration and integrate syntax checking, making the workflow compatible with incremental development rather than wholesale replacement.
\end{itemize}

\section{Method}\label{sec:method}


We propose a novel framework, referred to as \app, as illustrated in Fig.~\ref{fig:method}. Our approach integrates a code knowledge graph with established software engineering design paradigms to emulate the cognitive processes of professional developers, enabling robust architectural-level reasoning and iterative adaptation to evolving requirements. In addition, we incorporate a rule-based refactoring mechanism to systematically restructure “spaghetti-like” code files, thereby substantially improving the maintainability and readability of the codebase.

Furthermore, we employ a sandboxed execution environment in conjunction with automated test scripts to verify the consistency between the generated code and user requirements. This design establishes a closed-loop feedback mechanism in which test outcomes continuously trigger updates to both the \kg. As a result, the framework promotes long-term consistency, correctness, and maintainability of the generated codebase.

As shown in Figure~\ref{fig:method}, the proposed framework consists of four sequential stages:

\begin{figure*}[t]
  \centering
  \includegraphics[width=0.75\textwidth]{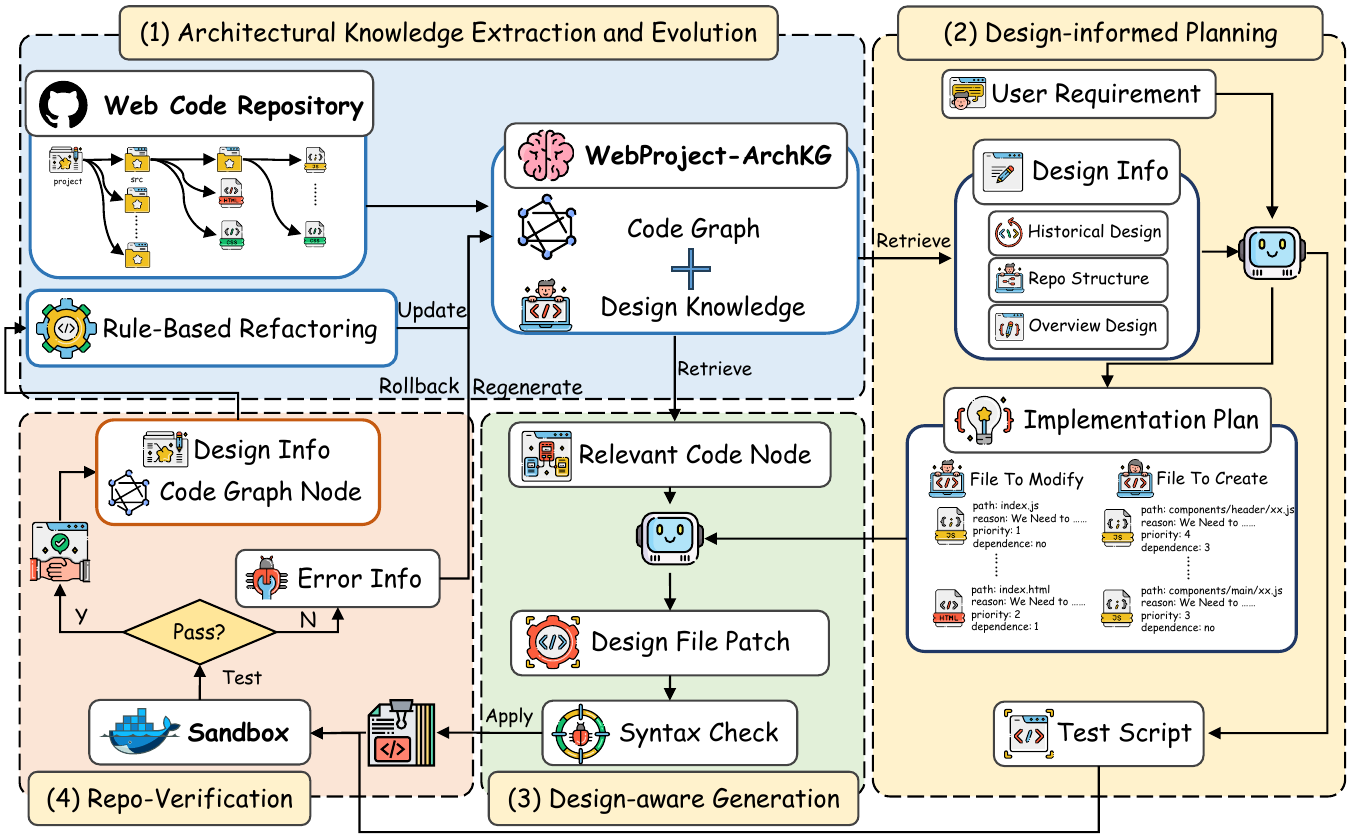}
  \caption{Overview of \app.}
  \label{fig:method}
\end{figure*}

\begin{enumerate}
  \item \textbf{Architectural Knowledge Extraction and Evolution:} 
    The initial code repository is first transformed into a knowledge graph that captures cross-file structural and semantic relationships. In parallel, a design blueprint is initialized by integrating historical design knowledge, the repository structure, and high-level architectural overview information. Prior to updating \kg, we apply rule-based refactoring to “spaghetti-like” code files to ensure that the information maintained in \kg is grounded in good maintainability and readability.

  \item \textbf{Design-informed Planning:} 
    Given the current user requirements, the framework retrieves relevant historical design information, repository structure, and architectural overview data from \kg. If feedback from previous iterations is available in \kg, these design and feedback signals are jointly provided to the LLM to generate an Implementation Plan and a corresponding set of Test Scripts.

  \item \textbf{Design-aware Generation:} 
    Guided by the Implementation Plan, the framework performs priority-aware execution by retrieving relevant contextual nodes from the code graph in \kg. Code changes are generated in the form of diff-based patches, enabling precise and minimal modifications to the existing codebase.

  \item \textbf{Repo-Verification:} 
    The generated patches are applied incrementally. The framework first performs AST-based static analysis to detect potential syntax errors; if any are found, the LLM is invoked to repair the patch in isolation. If the static checks pass, the patch is validated in a sandboxed environment. Upon successful validation, rule-based refactoring is applied to any newly introduced “spaghetti-like” code, followed by an update to \kg. If validation fails, a feedback mechanism is triggered to record the failure signals in \kg, which are then used to regenerate the Implementation Plan.
\end{enumerate}

\subsection{Architectural Knowledge Extraction and Evolution}

\begin{figure*}[t]
  \centering
  \includegraphics[width=0.85\textwidth]{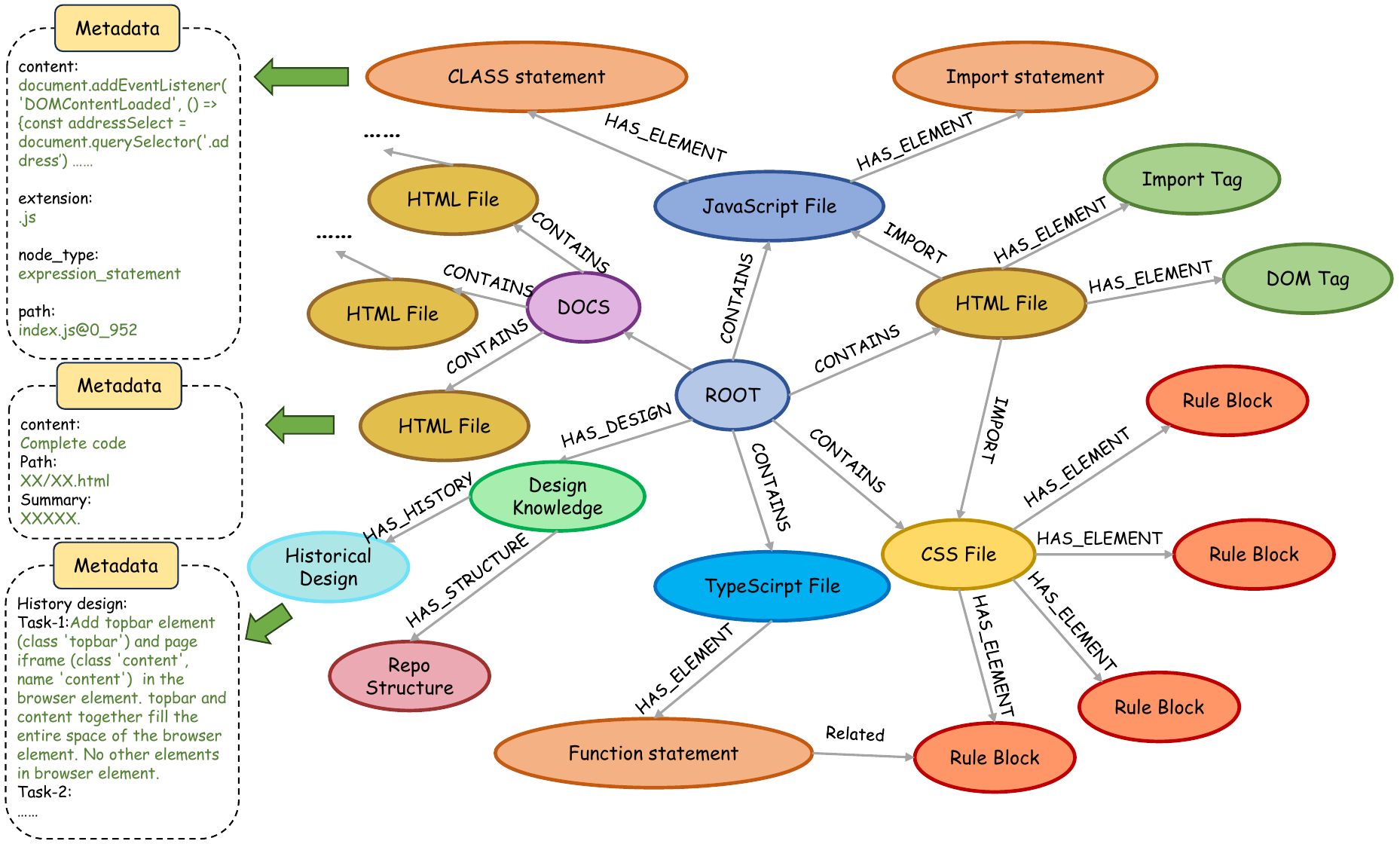}
  \caption{Schema of \kg.}
  \label{fig:code_graph}
\end{figure*}

In this stage, we focus on systematically extracting and continuously evolving architectural knowledge from the initialized front-end repository. By analyzing the repository, we construct the \kg, which provides an explicit and machine-interpretable, repository-level representation of file-level and component-level dependencies. It captures the semantic information of the code and explicitly models the connections between design knowledge and the code repository, as illustrated in Figure~\ref{fig:code_graph}. This representation forms the foundation of the architectural knowledge and is progressively refined throughout subsequent iterations.

\begin{figure}[htbp]
  \centering
  \includegraphics[width=1.0\columnwidth]{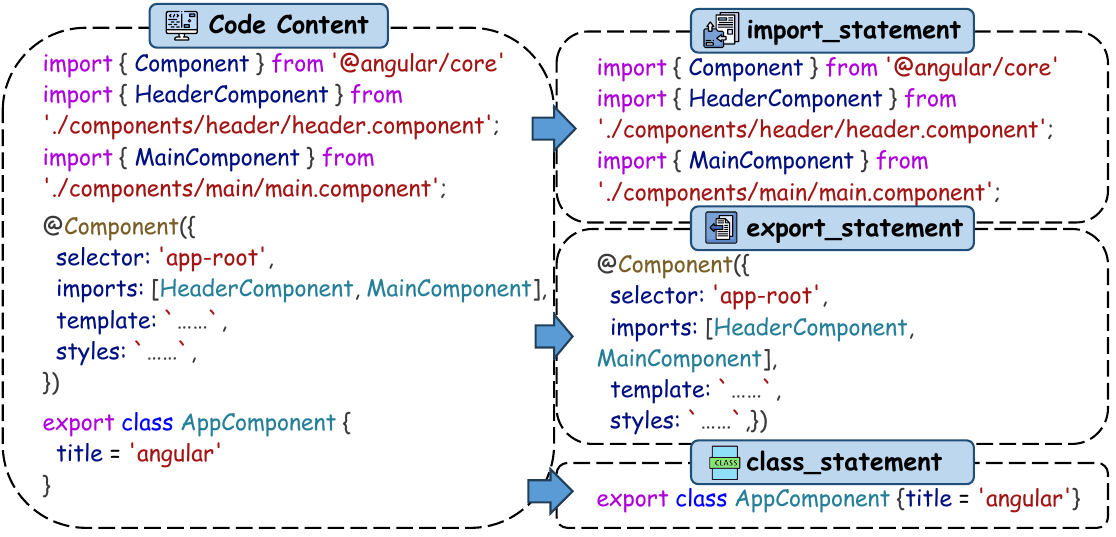}
  \caption{An Example of Code Content Segmentation into Blocks.}
  \label{fig:code_segments}
\end{figure}

The extraction and evolution process consists of the following steps:

\begin{itemize}
    \item Extract architectural primitives by parsing source files into abstract syntax trees and segmenting them into block-level units, as shown in Figure~\ref{fig:code_segments}.
    \item Infer and refine cross-file architectural relationships by analyzing inter-block references and dependency patterns across the repository.
    \item Evolve the architectural representation by refactoring heterogeneous code structures, extracting embedded code written in different languages into separate files, and reconstructing dependencies using the same extraction pipeline.
    \item Initialize the design knowledge in \kg, which represents the current state of the codebase as well as its complete design information and architectural memory.
\end{itemize}

\textbf{Architectural Primitive Extraction.}  
As illustrated in Figure~\ref{fig:code_segments}, each source file is parsed using abstract syntax tree analysis. We employ Tree-sitter to extract descendant block-level nodes from the syntax tree and treat them as architectural primitives. These primitives constitute the fundamental units of architectural knowledge and enable fine-grained modeling of component structure and behavior in downstream stages.

\textbf{Architectural Relationship Inference.}  
Building upon the extracted primitives, we infer architectural relationships at both coarse-grained and fine-grained levels. At the coarse-grained level, file-level dependencies are identified by detecting explicit inclusion and import signals, such as \texttt{<script>} tags in HTML files and \texttt{import} statements in JavaScript or TypeScript files. 

At the fine-grained level, we analyze inter-block references to identify selectors, identifiers, and referenced elements, and align them with symbols defined in related files. When a correspondence is established, a precise architectural linkage is recorded in the \kg, progressively refining the architectural representation.

This process links cross-file and interrelated code components within \kg, constructing a component-centric Code Graph that fully captures contextual information, thereby enabling precise understanding and reasoning over the code.

\textbf{Architectural Evolution via Refactoring.}  
To enable the continuous evolution of extracted architectural knowledge, the system scans both during \kg initialization and upon receiving modified code files to detect embedded heterogeneous code fragments, such as JavaScript within HTML files or HTML snippets within TypeScript files. Using a rule-based extraction strategy, these fragments are separated into new files, while the original files are updated to maintain consistency.

This refactoring step evolves the architectural representation by improving modular boundaries, reducing file complexity, and preserving consistency within the dependency structure encoded in the knowledge graph.

\textbf{Architectural Memory Initialization.}  
Finally, the framework clears the Historical Design within the Design Knowledge in \kg, while the Repository Structure is constructed as a tree representation derived from the initial directory layout. The Architectural Overview is initialized as empty, allowing it to be gradually populated and updated throughout successive design and implementation cycles.

\subsection{Design-informed Planning}

In this stage, we transform user requirements into a system-level high-level design plan that emulates a realistic software development workflow. The primary objective of this plan is to guide the LLM, acting as a developer, to reason over the repository structure, the architectural overview, and the historical design context, and to synthesize a structured modification strategy. This strategy explicitly specifies which files should be modified or created, the recommended order of execution, and the existing artifacts that should be referenced. By providing sufficient and accurate contextual grounding, the plan supports subsequent code generation and effectively mitigates the risk of hallucinated or inconsistent outputs.

In parallel, the framework automatically derives a set of end-to-end test scripts aligned with the intended functionality of the design plan. These tests serve as an executable specification of functional correctness and provide a verification mechanism for the generated implementation.

Specifically, the process can be divided into the following two steps:

\textbf{Design Knowledge-Guided Implementation Plan Generation.} 
In this step, \app automatically retrieves the corresponding Design Info from \kg based on the user requirements and current state. Specifically, the system queries the latest state of \kg and retrieves the stored records of the most recent Historical Design. It then scans the directory structure in \kg, organizing it into a tree representation presented as the Repo Structure. Finally, the system extracts the file-level Overview Design from each file node, organizes it in an ordered manner, and returns it as complete Design Info to guide the LLM in generating the Implementation Plan.

\textbf{Design Knowledge-Guided Test Script Generation.} 
The generation of test script is performed synchronously within the aforementioned workflow. A sample is provided here to illustrate the required output format. The reason for not generating them separately is that doing so may prevent the model from accurately understanding its most recent actions, potentially resulting in test scripts that do not align with the task requirements.

\subsection{Design-aware Generation}

In the design-aware generation phase, the framework follows the Implementation Plan produced in the previous stage and executes code generation in a priority-aware manner. For each planned operation, the framework references the contextual information of dependent files and retrieves pre-established dependencies from the \kg. Neighboring nodes related to the target files are automatically included in the prompt to construct a well-founded contextual representation, preventing the model from generating fabricated outputs due to incomplete knowledge of repository-level dependencies.

Based on this enriched contextual grounding, the system generates a corresponding set of \emph{Design File Patches} for each high-level design action, ensuring that the implementation remains aligned with the intended architectural and functional objectives.

\begin{figure*}[t]
  \centering
  \includegraphics[width=0.90\textwidth]{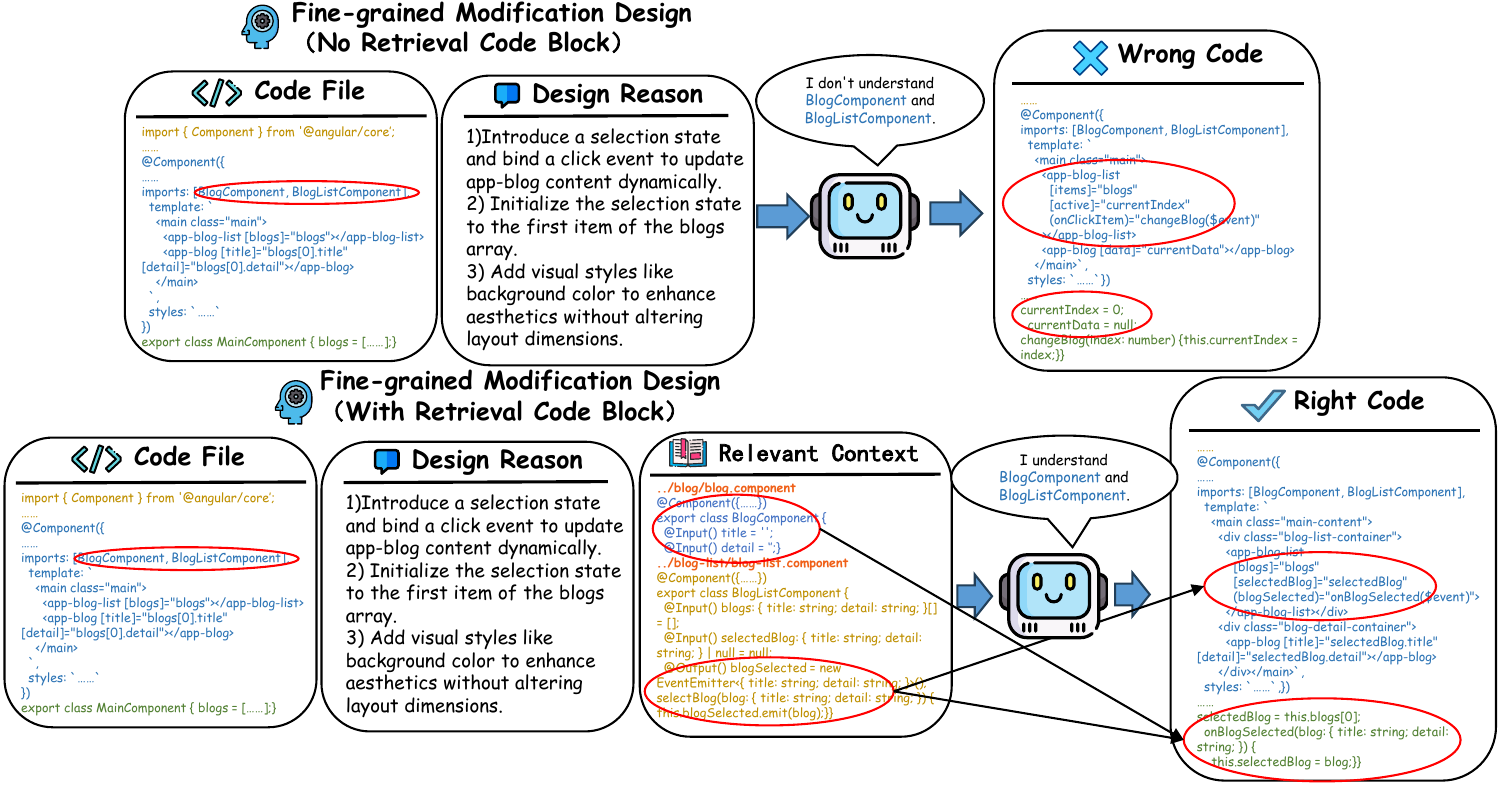}
  \caption{Impact of \kg--Based Retrieval on Generated Code Correctness.}
  \label{fig:case_code_retrive}
\end{figure*}

\textbf{Context Construction Based on the \kg.} 
In our approach, the \kg is constructed in advance and serves as a structured source of contextual information during code generation. When generating code, the system automatically retrieves nodes connected by relevant edges in the graph and aggregates their associated information as contextual input. As illustrated in Figure ~\ref{fig:case_code_retrive}, the retrieval of semantically and structurally related code context enables the model to reason about which existing artifacts should be referenced, rather than fabricating nonexistent functions or interfaces, thereby substantially reducing the risk of hallucinated outputs.

\textbf{Unified Patch-Based Output Format.}
We adopt a patch-based output representation rather than generating complete source files. To the best of our knowledge, most existing studies \cite{codes, repocoder} focus on full-file code generation, which significantly increases token consumption and introduces the risk of inadvertently overwriting correct code or hallucinating the removal of valid implementations.

To address these issues, the system uses a unified diff format to represent code modifications in a minimal and localized manner. Furthermore, since LLMs often struggle to accurately track line numbers and boundaries, a rule-based mechanism is employed to automatically correct erroneous diff outputs, including incorrect start and end line indices. This mechanism ensures that the generated patches can be reliably and consistently applied to the target repository.
\subsection{Repo-Verification}
At this stage, the system directly applies the diff-based patch generated in the previous stage to the target codebase. It then performs an abstract syntax tree analysis to detect potential syntax errors. If no errors are found, execution proceeds; if errors are detected, the LLM is invoked to repair the individually modified files. If repairs fail more than twice, the LLM generates the complete current code guided by the patch to ensure that the intended changes can be correctly applied. After successful repair, the updated codebase is deployed in a Docker-based sandbox environment for verification, where the test scripts produced during the high-level system design phase are executed to ensure that the generated implementation meets user requirements.

Upon successful validation, the system performs a refactoring scan of the applied code, and both the refactored and unrefactored files are fed into \kg for node expansion and dependency construction. Meanwhile, a LLM summarizes the newly implemented functionalities from this deployment, which are incorporated into the Overview Design within the Design Knowledge. The historical requirements are recorded and added to the Historical Design, updating \kg and the Repository Structure accordingly.

\section{Evaluation}\label{sec:evaluation}

To evaluate the effectiveness of \app, we conduct extensive experiments addressing the following research questions:

\begin{itemize}
    \item \textbf{RQ1 (Effectiveness):} How does \app perform in terms of task completion and reliability for incremental front-end repository-level code generation, compared to the Web-Agent baseline?

    \item \textbf{RQ2 (Ablation Study):} How do the individual components of the framework contribute to its overall effectiveness?

    \item \textbf{RQ3 (Robustness and Maintainability):}
    How do the design paradigm and refactoring strategies in \app affect failure modes and code maintainability?

    \item \textbf{RQ4 (Cost Efficiency):} How do representative general-purpose coding agents compare in task completion, token consumption, and cost-effectiveness?
\end{itemize}

\subsection{Experimental Setup}
\noindent\textbf{Benchmark.} 
We adopt Web-Bench~\cite{webbench}, the first benchmark specifically designed for natural-language-to-repository-level frontend code generation. The benchmark comprises 50 projects, each with 20 sequential tasks that introduce new requirements depending on the repository state produced by prior tasks---closely reflecting real-world incremental frontend development. Projects span modern frameworks including Vue, Angular, and Tailwind.

\noindent\textbf{Baseline.} 
We adopt Web-Agent~\cite{webbench} as our primary baseline---the only publicly evaluated agent designed for repository-level front-end code generation, leveraging file manipulation tools and a RAG system. In addition, we compare against four widely adopted general-purpose coding agents---Codex CLI, Claude Code, OpenHands, and SWE-Agent---to assess how representative tools perform on front-end tasks without design knowledge. While these agents have demonstrated strong results on general software engineering benchmarks such as SWE-bench, they were not designed for the incremental, heterogeneous, multi-framework nature of front-end development. Most existing studies~\cite{agentless, repocoder, sweagent, chainagent} target non-front-end tasks, while front-end approaches remain limited to single-component UI generation or design-to-UI conversion~\cite{laurencon2024unlocking, designcoder, UICopilot}, operating at shallow abstraction levels that fail to capture real-world repository-level workflows.

\noindent\textbf{Evaluated Models.}
For RQ1--RQ3, we follow the original Web-Bench study~\cite{webbench} and evaluate the same nine foundation models spanning the Gemini, Claude, Qwen, OpenAI, and DeepSeek series: Gemini 2.5 Pro, Claude 4 Sonnet, GPT-4.1, GPT-4o, O4-Mini, Qwen-Max, Qwen-Plus, DeepSeek-V3, and DeepSeek-R1. For RQ4, we extend the evaluation to two additional model families---DeepSeek V4 (Pro and Flash) and Qwen 3.5 (Plus and Flash)---paired with general-purpose coding agents and \app under identical benchmark conditions. All models are evaluated on the same set of projects for fair comparison. This setup provides a stronger stress test: the RQ4 models and general-purpose agents represent more recent and capable configurations, making \app's consistent outperformance across all of them a particularly stringent validation of its design-knowledge-driven approach.

\subsection{RQ1: Effectiveness}
\noindent\textbf{Design.} 
We compare \app against Web-Agent across the same nine foundation models originally evaluated in the Web-Bench study~\cite{webbench} on the full benchmark. For each configuration, we report Pass@1 (fraction of tasks completed on the first attempt), Pass@2 (fraction completed within two attempts). All metrics are computed across the complete set of 50 projects.

\begin{table*}[t]
\centering
\small
\caption{Performance comparison between Web-Agent and \app across foundation models. 
Overall metrics are computed across all evaluated projects.
Improvements are reported in absolute percentage points relative to Web-Agent.}
\label{tab:main_results}
\begin{tabular}{@{}lcccc@{}}
\toprule
\multirow{2}{*}{\textbf{Model}} & \multicolumn{2}{c}{\textbf{\app (Ours)}} & \multicolumn{2}{c}{\textbf{Web-Agent}} \\
\cmidrule(lr){2-3} \cmidrule(lr){4-5}
& \textbf{Pass@1} & \textbf{Pass@2} & \textbf{Pass@1} & \textbf{Pass@2} \\
& \textbf{(\%)} $\uparrow$ & \textbf{(\%)} $\uparrow$ & \textbf{(\%)} $\uparrow$ & \textbf{(\%)} $\uparrow$ \\
\midrule
Gemini 2.5 Pro  & \textbf{36.20} & \textbf{49.40} & 26.40 & 42.40 \\
Claude 4 Sonnet & \textbf{34.90} & \textbf{48.10} & 24.30 & 39.70 \\
GPT-4.1         & \textbf{34.20} & \textbf{37.10} & 21.09 & 25.11 \\
Qwen-Max        & \textbf{27.90} & \textbf{31.20} & 15.87 & 19.02 \\
DeepSeek-V3     & \textbf{23.20} & \textbf{36.60} & 17.07 & 23.59 \\
DeepSeek-R1     & \textbf{23.20} & \textbf{39.00} & 17.10 & 23.20 \\
GPT-4o          & \textbf{21.30} & \textbf{28.80} & 17.17 & 23.80 \\
O4-Mini         & \textbf{17.90} & \textbf{28.50} & 13.26 & 22.93 \\
Qwen-Plus       & \textbf{16.30} & \textbf{22.10} & 10.98 & 15.11 \\
\midrule
\textbf{Avg. $\Delta$} & \textbf{+7.98} & \textbf{+9.55} & -- & -- \\
\bottomrule
\end{tabular}
\end{table*}

\begin{figure}[htbp]
  \centering
  \includegraphics[width=1.0\columnwidth]{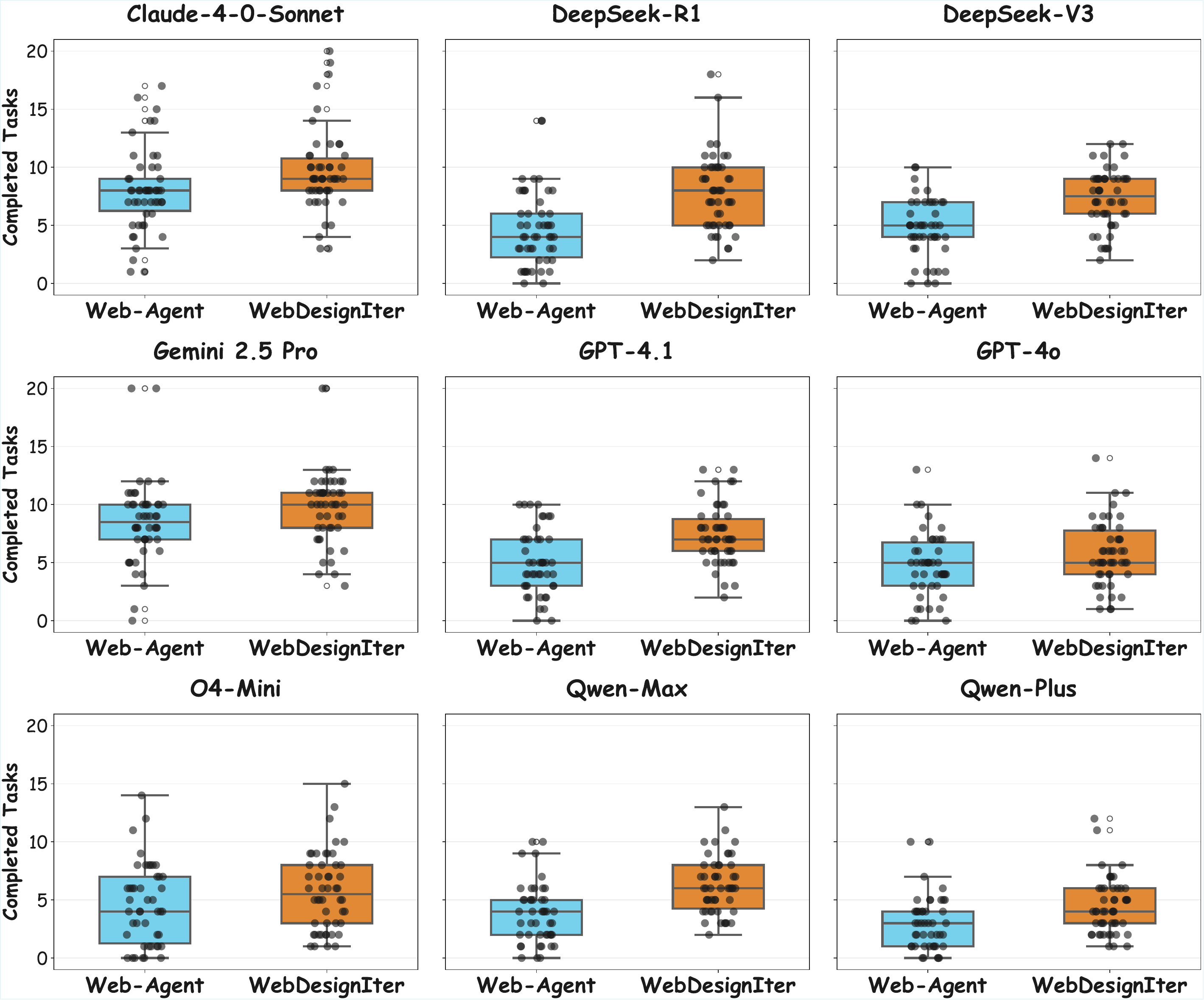}
  \caption{Comparison of task completion per project between Web-Agent and our method (\app) across multiple models. Each boxplot summarizes the distribution of completed tasks (pass@2 × 20) across all projects within a model, with Web-Agent depicted in blue and \app in orange. Individual project results are overlaid as jittered points to illustrate variability.}
  \label{fig:boxplot}
\end{figure}

To complement the aggregate metrics, Figure~\ref{fig:boxplot} visualizes the per-project task completion distribution via boxplots, revealing variability and outlier behavior across projects.

\noindent\textbf{Result.}
As shown in Table~\ref{tab:main_results}, \app consistently outperforms Web-Agent across all evaluated models. The best Web-Agent configuration---Gemini 2.5 Pro---achieves 26.40\% Pass@1, 42.40\% Pass@2. With the same model, \app attains 36.20\% Pass@1, 49.40\% Pass@2. Averaged across all nine models, \app delivers gains of +7.98\% in Pass@1, +9.55\% in Pass@2. Improvements are consistent across both commercial models (Gemini 2.5 Pro, GPT-4.1, Claude 4 Sonnet) and open-source models (DeepSeek-R1, Qwen-Max). The boxplots in Fig.~\ref{fig:boxplot} further confirm that \app shifts the overall distribution of completed tasks upward relative to Web-Agent.

\begin{samepage}
\begin{mdframed}[backgroundcolor=gray!15,linecolor=gray,roundcorner=5pt]
\textbf{Finding 1:} \app consistently outperforms Web-Agent across all evaluated models, achieving an average improvement of +9.55\% in Pass@2 and demonstrating particular strength on complex tasks within Web-Bench.
\end{mdframed}
\end{samepage}

\subsection{RQ2: Ablation Study}
To systematically analyze the contribution of each component, we conduct ablation experiments using Claude 4 Sonnet---the best-performing model in RQ1---by progressively disabling individual modules from the full \app framework. All variants are evaluated under identical protocols.

\noindent\textbf{Design.} 
We examine four core components:
(1) the Design module, which injects high-level architectural constraints and global design intent;
(2) the Code Graph retrieval module, which models repository-level structure and cross-file dependencies;
(3) the file patching mechanism, which performs targeted incremental updates rather than full-file rewriting; and
(4) the sandbox-driven iterative module, which refines outputs based on automated execution and test feedback.

\begin{table*}[t]
\centering
\small
\caption{Ablation study of our framework on the Web-Bench~\cite{webbench} benchmark. 
Each variant disables one key module to quantify its contribution. 
Improvements are reported in absolute percentage points relative to the Full System.}
\label{tab:ablation}
\setlength{\tabcolsep}{6pt}
\begin{tabular}{lcccccc}
\toprule
\multirow{2}{*}{\textbf{Variant}} 
& \multicolumn{4}{c}{\textbf{Modules}} 
& \multicolumn{2}{c}{\textbf{Overall (\%)}} \\
\cmidrule(lr){2-5} \cmidrule(lr){6-7}
& \textbf{Design} 
& \textbf{Code Graph} 
& \textbf{Patch} 
& \textbf{Sandbox} 
& \textbf{Pass@1 $\uparrow$} 
& \textbf{Pass@2 $\uparrow$} \\
\midrule

\textbf{Full System} 
& \checkmark & \checkmark & \checkmark & \checkmark 
& \textbf{34.90} 
& \textbf{48.10} \\

\midrule
\textbf{w/o Design} 
& $\times$ & \checkmark & \checkmark & \checkmark 
& 23.50 \, (-11.40) 
& 42.10 \, (-6.00) \\

\textbf{w/o Code Graph} 
& \checkmark & $\times$ & \checkmark & \checkmark 
& 26.50 \, (-8.40) 
& 41.10 \, (-7.00) \\

\textbf{w/o Patch} 
& \checkmark & \checkmark & $\times$ & \checkmark 
& 31.40 \, (-3.50) 
& 42.10 \, (-6.00) \\

\textbf{w/o Sandbox} 
& \checkmark & \checkmark & \checkmark & $\times$ 
& 31.60 \, (-3.30) 
& 43.20 \, (-4.90) \\

\bottomrule
\end{tabular}
\end{table*}

\noindent\textbf{Result.}

\noindent\textbf{w/o Sandbox.}
Removing the sandbox yields the best performance among all ablation variants, achieving the highest Pass@1 and Pass@2. This suggests that the sandbox primarily serves as a verification and feedback mechanism rather than a direct contributor to generation quality. Its role is to execute code and surface runtime errors for iterative refinement, not to participate in code synthesis itself.

\begin{figure}[htbp]
  \centering
  \includegraphics[width=0.85\columnwidth]{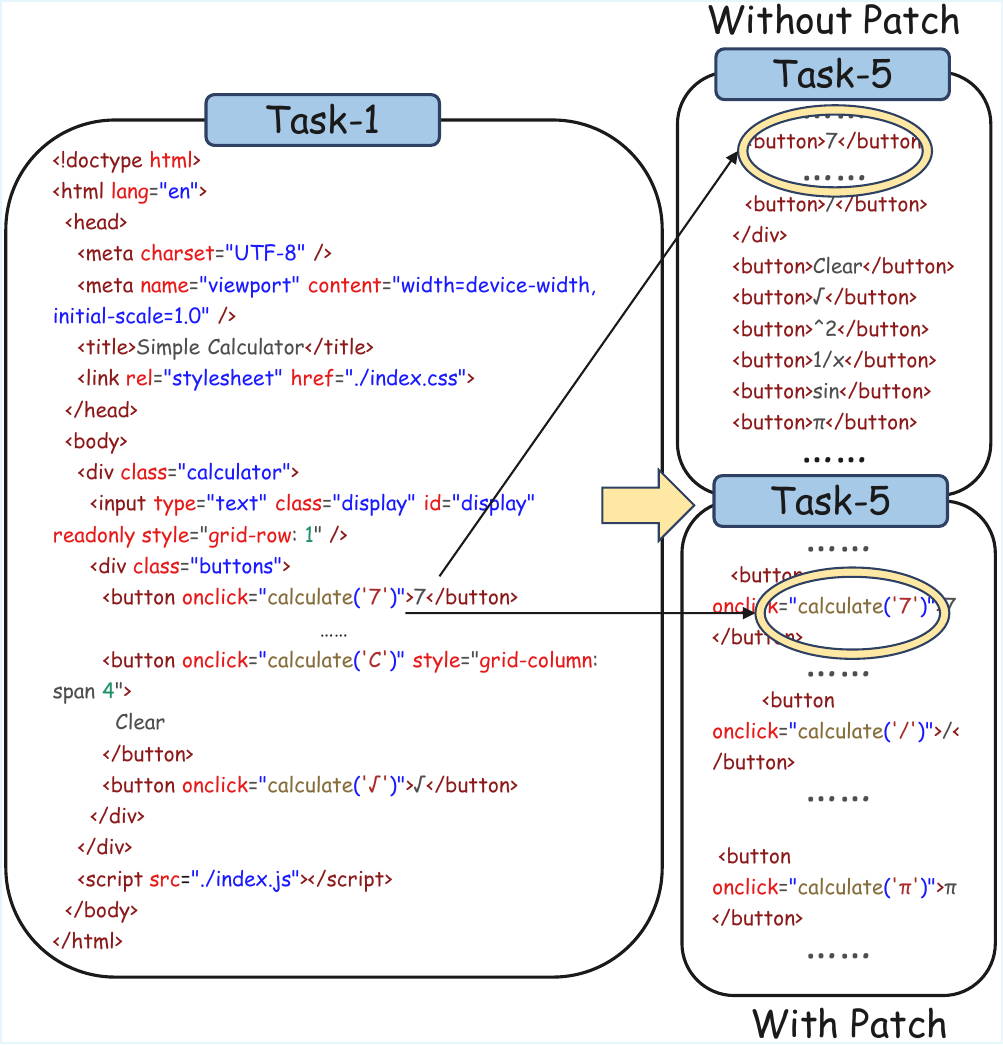}
  \caption{A Case Study on Patch Strategy and Code Generation.}
  \label{fig:case_patch}
\end{figure}

\noindent\textbf{w/o Patch.}
Removing the patch mechanism results in a moderate performance drop (Pass@1: ---3.50\%, Pass@2: ---6.00\%), ranking second among variants. The sequential nature of the benchmark---where each task depends on prior state---amplifies the impact of unnecessary rewrites. As illustrated in Fig.~\ref{fig:case_patch}, without patching, the model at Task-5 overwrites functions correctly implemented at Task-1, causing cascading regressions. The patch mechanism prevents this by limiting updates to only the most relevant code segments, preserving unchanged components across tasks.

\noindent\textbf{w/o Code Graph.}
Removing Code Graph causes a substantial degradation (Pass@1: ---8.40\%, Pass@2: ---7.00\%), confirming its critical role. The Code Graph provides the model with cross-file dependency context; without it, the model struggles to trace dependencies across files, increasing hallucinated or inconsistent generation. Fig.~\ref{fig:case_code_retrive} further illustrates how Code Graph mitigates cross-file misinterpretations.

\noindent\textbf{w/o Design.}
Removing the Design module produces the largest drop (Pass@1: ---11.40\%, Pass@2: ---6.00\%). The Design module encapsulates system-level architectural knowledge---file summaries, module responsibilities, and structural constraints. Without this guidance, the model cannot efficiently localize relevant files or understand functional roles, increasing uncertainty in modification targets. Moreover, historical design knowledge provides persistent guidance across tasks, reducing functional regression.
\begin{samepage}
\begin{mdframed}[backgroundcolor=gray!15,linecolor=gray,roundcorner=5pt]
\textbf{Finding 2:} 
Every component of \app---Design, Code Graph, Patch, and Sandbox---contributes measurably to overall performance. The Design module and Code Graph are the most impactful, while Patch and Sandbox provide complementary benefits in consistency and verification.
\end{mdframed}
\end{samepage}
\subsection{RQ3: Robustness and Maintainability}
\noindent\textbf{Design.}
\app leverages explicit design modeling and global context to reduce failures from cross-file errors, API dependencies, and regressions. It also automatically refactors redundant content, promoting concise and maintainable outputs. To quantify this, we introduce \emph{average file length}---the mean size of output files in each project's final task---as a proxy for code locality and redundancy.

We further categorize unresolved errors (failures persisting after two attempts) into five types: \textbf{Style} (visual non-compliance), \textbf{Cross-File} (incorrect cross-file references), \textbf{Dependency} (wrong API or package usage), \textbf{Function} (logical errors without the above patterns), and \textbf{Regression} (previously passing tasks that break after subsequent modifications).

\begin{figure}[htbp]
  \centering
  \includegraphics[width=0.6\columnwidth]{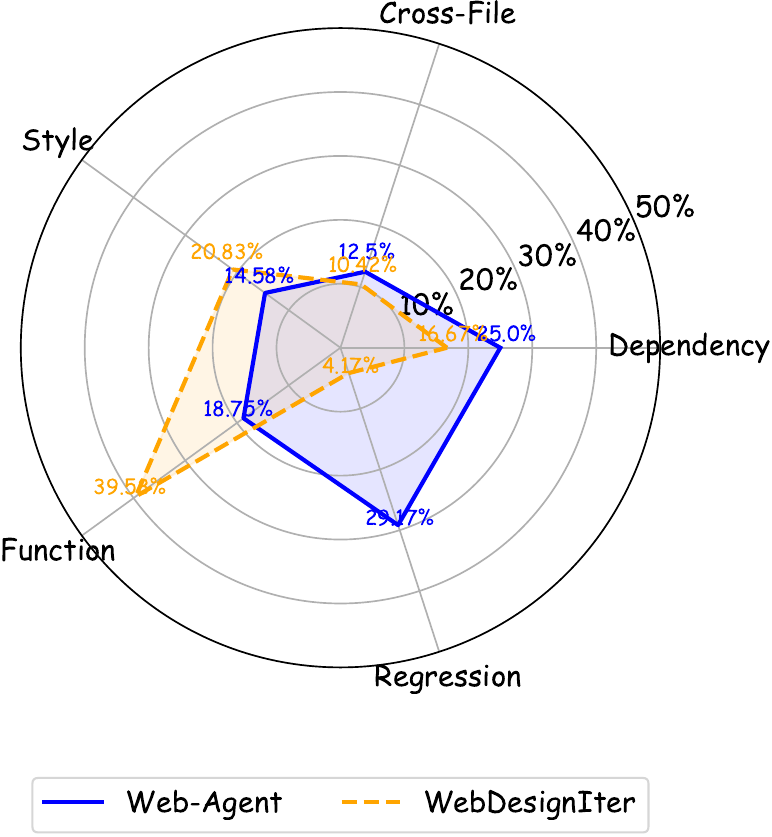}
  \caption{Radar chart of error type distributions for Web-Agent and the \app method.}
  \label{fig:error_type}
\end{figure}

\noindent\textbf{Results.}
Fig.~\ref{fig:error_type} compares error-type distributions between Web-Agent and \app. \app substantially reduces \emph{Regression} errors (29.17\% $\rightarrow$ 4.17\%), \emph{Dependency} errors (25.00\% $\rightarrow$ 16.67\%), and \emph{Cross-File} errors (12.50\% $\rightarrow$ 10.42\%). This reduction aligns with the patch mechanism's ability to minimize collateral modifications (Fig.~\ref{fig:case_patch}) and the design knowledge's cross-file context.

Conversely, \emph{Function} errors increase (18.75\% $\rightarrow$ 39.58\%) and \emph{Style} errors increase moderately (14.58\% $\rightarrow$ 20.83\%). This shift does not indicate degraded performance; rather, by eliminating regression and dependency failures, \app allows the system to reach deeper stages of functional construction where more complex implementation- and logic-level errors are exposed. These error types fundamentally depend on the base model's reasoning capability, not on architectural awareness.

Consistent with these findings, the average file length metric (Table~\ref{tab:rq3}) shows that \app produces more localized and concise code, confirming effective refactoring of redundant content.

\begin{table}[t]
\centering
\small
\caption{Comparison of code modularity between Web-Agent and \app.}
\label{tab:rq3}
\setlength{\tabcolsep}{8pt}
\begin{tabular}{lcc}
\toprule
\textbf{Method} 
& \textbf{Avg. File Length} 
& \textbf{Avg. Modified Files} \\
\midrule

Web-Agent           
& 49.41  
& 515 \\

\textbf{\app} 
& \textbf{35.38 (↓28.4\%)}  
& \textbf{612 (↑18.8\%)} \\

\bottomrule
\end{tabular}
\end{table}

\begin{samepage}
\begin{mdframed}[backgroundcolor=gray!15,linecolor=gray,roundcorner=5pt]
\textbf{Finding 3:} 
\app shifts the dominant failure mode from regression and dependency errors toward higher-level functional and logical challenges, while producing more concise code. This indicates that the primary performance bottleneck moves from architectural consistency to the base model's reasoning capabilities.
\end{mdframed}
\end{samepage}
\subsection{RQ4: Cost Efficiency}
\noindent\textbf{Motivation.}
While RQ1--RQ3 establish \app's effectiveness against Web-Agent, a natural question is how general-purpose coding agents perform on the same benchmark and at what token cost.

\noindent\textbf{Design.}
We evaluate \app alongside Claude Code, OpenHands, SWE-Agent, and Codex CLI on the full Web-Bench benchmark under identical conditions. For Claude Code, OpenHands, SWE-Agent, and \app, we report results with four models spanning two families: DeepSeek V4 (Pro and Flash) and Qwen 3.5 (Plus and Flash). Codex CLI is evaluated with GPT-5.5 and GPT-5.2. For each configuration, we report Pass@1, Pass@2, and input/output token consumption.

\begin{table}[t]
\centering
\scriptsize
\caption{Comparison of \app with general-purpose coding agents on Web-Bench.
All agents are evaluated on the same 50-project benchmark under identical conditions.
Token consumption is reported in thousands (K).
\textbf{Bold} marks the best result in each column.}
\label{tab:coding_agents}
\setlength{\tabcolsep}{3pt}
\begin{tabular}{@{}llcccc@{}}
\toprule
\textbf{Method} & \textbf{Model} & \textbf{Pass@1} & \textbf{Pass@2} & \textbf{In Tok.} & \textbf{Out Tok.} \\
& & \textbf{(\%)} $\uparrow$ & \textbf{(\%)} $\uparrow$ & \textbf{(K)} $\downarrow$ & \textbf{(K)} $\downarrow$ \\
\midrule
\multirow{4}{*}{\textbf{\app}} 
& DeepSeek V4 Pro      & \textbf{33.53} & \textbf{53.14} & \textbf{533} & 317 \\
& DeepSeek V4 Flash    & \textbf{29.51} & \textbf{47.65} & \textbf{525} & 284 \\
& Qwen 3.5 Plus        & \textbf{32.16} & \textbf{50.29} & \textbf{523} & 306 \\
& Qwen 3.5 Flash       & \textbf{26.08} & \textbf{44.31} & \textbf{528} & 291 \\
\midrule
\multirow{4}{*}{Claude Code} 
& DeepSeek V4 Pro      & 21.27 & 35.59 & 13,949 & 2,218 \\
& DeepSeek V4 Flash    & 18.43 & 36.76 & 15,491 & 3,074 \\
& Qwen 3.5 Plus        & 19.51 & 37.94 & 14,230 & 2,829 \\
& Qwen 3.5 Flash       & 18.43 & 36.76 & 14,203 & 2,847 \\
\midrule
\multirow{4}{*}{OpenHands}   
& DeepSeek V4 Pro      & 26.70 & 29.60 & 810   & 984 \\
& DeepSeek V4 Flash    & 23.20 & 25.80 & 826   & 1,014 \\
& Qwen 3.5 Plus        & 24.70 & 27.30 & 813   & 996 \\
& Qwen 3.5 Flash       & 23.10 & 25.70 & 805   & 998 \\
\midrule
\multirow{4}{*}{SWE-Agent}   
& DeepSeek V4 Pro      & 20.80 & 25.20 & 613   & 605 \\
& DeepSeek V4 Flash    & 17.90 & 22.00 & 633   & 625 \\
& Qwen 3.5 Plus        & 18.80 & 22.90 & 609   & 591 \\
& Qwen 3.5 Flash       & 17.90 & 22.00 & 604   & 597 \\
\midrule
\multirow{2}{*}{Codex CLI}   
& GPT-5.5              & 8.50  & 14.00 & 16,120 & 2,347 \\
& GPT-5.2              & 5.00  & 9.00  & 14,890 & 2,189 \\
\bottomrule
\end{tabular}
\end{table}

\noindent\textbf{Result.}
As shown in Table~\ref{tab:coding_agents}, \app consistently outperforms all general-purpose agents across every model configuration while consuming substantially fewer tokens. With DeepSeek V4 Pro, \app achieves 33.53\% Pass@1 and 53.14\% Pass@2 using 533K input tokens---approximately 26$\times$ fewer than Claude Code (13,949K). This pattern holds across all four models: \app attains 26.08\%--33.53\% Pass@1 with 523--533K input tokens, whereas Claude Code requires 14--15M tokens to reach only 18.43\%--21.27\%. OpenHands and SWE-Agent, though more token-efficient than Claude Code (600--1,000K), achieve only 16.50\%--26.00\% Pass@1. Codex CLI trails at 5.00\%--8.50\% Pass@1 despite consuming over 15M input tokens.

This efficiency gap reflects fundamentally different design philosophies. General-purpose agents rely on iterative trial-and-error within an agentic loop, incurring heavy overhead from repeated file reading, failed attempts, and context re-evaluation---a cost that scales poorly with task complexity. \app's design-knowledge-driven workflow instead front-loads architectural knowledge, enabling targeted generation with minimal retry. The result is a qualitative shift in the cost--performance trade-off: \app simultaneously achieves the highest success rates and the lowest token cost, a combination that no general-purpose agent attains regardless of model capability.
\begin{samepage}
\begin{mdframed}[backgroundcolor=gray!15,linecolor=gray,roundcorner=5pt]
\textbf{Finding 4:} 
\app achieves the highest Pass@1 and Pass@2 across all evaluated configurations while consuming 25--30$\times$ fewer input tokens than agentic-loop baselines, demonstrating that design knowledge delivers a fundamentally superior cost--performance trade-off beyond what model scaling alone can provide.
\end{mdframed}
\end{samepage}
\section{Threats to Validity}
\textbf{Internal Validity.}
A primary threat to internal validity arises from the retrieval of relevant code context. Traditional retrieval-augmented generation (RAG) methods typically rely on similarity-based matching to identify contextual nodes. While often effective, these approaches can become unreliable when user requirements are implicit or poorly specified. In such cases, semantically similar code snippets may be absent, or the current task may not relate to any existing code at all. Consequently, similarity-based retrieval may return irrelevant or misleading context, increasing the likelihood of subsequent generation errors. This challenge highlights the need for more deterministic and structured context retrieval mechanisms that can reliably capture task-relevant information even under ambiguous or sparse user input.

In contrast, our approach employs a rule-based name-matching strategy grounded on a pre-constructed code graph. By explicitly modeling file- and symbol-level relationships and performing deterministic matching, the system systematically collects highly relevant contextual information. This design significantly reduces context misalignment and retrieval noise, mitigating internal validity threats. Moreover, leveraging the structured code graph enables precise cross-file reasoning, supporting reliable and consistent code generation in complex, repository-level front-end projects.

\textbf{External Validity.}
A major threat to external validity arises from the heterogeneity of pretraining corpora across different LLMs. Models trained on diverse data distributions may exhibit varying proficiency across programming frameworks, libraries, and coding styles~\cite{starcoder, codestyle}, which can lead to inconsistent code generation performance across repositories. Consequently, the observed effectiveness of \app could partially reflect the inherent strengths or weaknesses of the backbone model rather than the method itself. To mitigate this, we evaluate \app across a broad spectrum of models—including open-source and commercial variants with distinct training pipelines and dataset compositions—reducing model-specific bias and enhancing the robustness and generalizability of our conclusions for front-end repository-level tasks.
\section{Related Work}
\subsection{Repository-Level Code Generation}
LLM-based code generation has progressed from function-level benchmarks such as HumanEval~\cite{humaneval} to class-level tasks like ClassEval~\cite{classeval}, and further to repository-level challenges exemplified by SWE-Bench~\cite{sweagent}, which evaluates cross-file modifications and execution-based validation in real-world projects.

Recent approaches incorporate structured context and automated workflows. Tao et al.~\cite{tao2025code} model inter-file dependencies via code graphs for back-end generation. SWE-Agent~\cite{sweagent} enables autonomous agent-based navigation and multi-step reasoning. ConTested~\cite{dcontested} employs test-driven verification loops, while Agentless~\cite{agentless} decomposes tasks into staged retrieval, localization, editing, and validation pipelines. Li et al.~\cite{li2024repomincoder} focus on mitigating information loss in prompt encoding for partially implemented repositories.

Despite strong performance, most existing work targets back-end or general-purpose scenarios and does not adequately address front-end repository-level generation involving modern frameworks, interactive logic, and interface structures.

\subsection{Front-End Repository-Level Code Generation}
Recent years have witnessed rapid progress in applying artificial intelligence to front-end~\cite{2019-Stocco-Proweb, 9438563}. To the best of our knowledge, most existing research primarily falls into the category of UI2Code, which focuses on translating visual user interfaces into corresponding front-end implementations. For example, Wu et al. propose LayoutCoder~\cite{wu2025mllm}, which leverages multimodal large language models (MLLMs) and integrates three key modules---element relationship construction, UI layout parsing, and layout-guided fusion---to guide the generation of front-end code from interface representations. Xu et al. introduce WebVIA~\cite{WebVIA}, a framework that employs vision-language models to analyze screenshots and incorporates a verification module and an interactive code generation module to produce front-end code with executable interactions. Gui et al. present UICopilot~\cite{UICopilot}, which adopts a hierarchical code generation strategy by first producing coarse-grained code and then refining it into fine-grained implementations, thereby mitigating the adverse effects of long and complex code contexts on MLLMs.

While these approaches demonstrate the effectiveness of LLMs in front-end code generation, they largely overlook the incremental nature of real-world software development and the role of design knowledge in guiding repository-level evolution and long-term maintainability.

\subsection{Application of Design Knowledge in Code Generation}
Design knowledge is essential in software engineering. Aleti et al. emphasize its role in managing complex systems while ensuring high-quality standards, and Santos et al.~\cite{santos2019understanding} and Winograd et al.~\cite{winograd1996bringing} highlight its importance for architectural integrity and development practices. GraphCoder~\cite{liu2024graphcoder} further shows that repository-level knowledge is crucial for code generation, and from a software engineering perspective, repository-level knowledge can also be considered design knowledge. Despite this, few studies explicitly incorporate design knowledge in either general or front-end repository-level code generation. Our approach leverages design knowledge to guide front-end generation, adopting an incremental, repository-level perspective and integrating automated refactoring to enhance code readability and maintainability, making the generation process more aligned with real-world development practices.

\section{CONCLUSION}
We presented \app, a design-knowledge-driven framework for incremental front-end repository-level code generation. \app maintains a persistent knowledge graph (\kg) that integrates repository structure with architectural and design principles, driving a two-stage pipeline of design-informed planning and design-aware generation. Experiments on Web-Bench demonstrate that \app consistently outperforms the Web-Agent baseline across nine foundation models, achieving an average improvement of 9.55\% in Pass@2, with the best configuration (Gemini 2.5 Pro) reaching 49.40\% Pass@2. Ablation studies confirm that design knowledge is the most impactful component, with its removal causing a 11.40\% drop in Pass@1. Compared to general-purpose coding agents, \app consumes 25--30$\times$ fewer input tokens while attaining the highest pass rates, demonstrating that design knowledge achieves fundamentally superior cost efficiency. These results establish that integrating design knowledge into the code generation pipeline is a promising direction for reliable, maintainable, and cost-effective front-end development.

\section{DATA AVAILABILITY}
To facilitate the replication study, we have released our data and code at:\url{https://github.com/SYSUSELab/WebDesignIter}. 

\bibliographystyle{IEEEtran}
\bibliography{bib}

\end{document}